\documentclass[10pt, conference, compsocconf]{IEEEtran}

\usepackage[noadjust]{cite}

\ifCLASSINFOpdf

\else

\fi

\let\oldbibliography\thebibliography
\renewcommand{\thebibliography}[1]{%
	\oldbibliography{#1}%
	\setlength{\itemsep}{0pt}%
}

\usepackage[figurename=Fig.]{caption}
\usepackage[tablename=Table.]{caption}
\usepackage{amsmath}
\usepackage{amsfonts}
\usepackage{commath}
\usepackage{algorithm}
\usepackage{float}
\usepackage[noend]{algpseudocode}
\usepackage{multicol}
\usepackage{graphicx}
\usepackage{subcaption}
\usepackage{mathtools}
\usepackage{filecontents}
\usepackage{balance}
\usepackage{multirow}
\usepackage{xcolor}

\makeatletter
\def\BState{\State\hskip-\ALG@thistlm}
\renewcommand{\ALG@beginalgorithmic}{\footnotesize}
\makeatother

\begin{document}

\title{\huge Task Runtime Prediction in Scientific Workflows Using an Online Incremental Learning Approach}

\author{\IEEEauthorblockN{Muhammad Hafizhuddin Hilman,
		Maria A. Rodriguez and
		Rajkumar Buyya}
	\IEEEauthorblockN{Cloud Computing and Distributed Systems (CLOUDS) Laboratory\\
		School of Computing and Information Systems \\The University of Melbourne, Australia}
	\IEEEauthorblockA{Email: hilmanm@student.unimelb.edu.au, \{marodriguez, rbuyya\}@unimelb.edu.au}}

\maketitle

\begin{abstract}
Many algorithms in workflow scheduling and resource provisioning rely on the performance estimation of tasks to produce a scheduling plan. A profiler that is capable of modeling the execution of tasks and predicting their runtime accurately, therefore, becomes an essential part of any Workflow Management System (WMS). With the emergence of multi-tenant Workflow as a Service (WaaS) platforms that use clouds for deploying scientific workflows, task runtime prediction becomes more challenging because it requires the processing of a significant amount of data in a near real-time scenario while dealing with the performance variability of cloud resources. Hence, relying on methods such as profiling tasks' execution data using basic statistical description (e.g., mean, standard deviation) or batch offline regression techniques to estimate the runtime may not be suitable for such environments. In this paper, we propose an online incremental learning approach to predict the runtime of tasks in scientific workflows in clouds. To improve the performance of the predictions, we harness fine-grained resources monitoring data in the form of time-series records of CPU utilization, memory usage, and I/O activities that are reflecting the unique characteristics of a task's execution. We compare our solution to a state-of-the-art approach that exploits the resources monitoring data based on regression machine learning technique. From our experiments, the proposed strategy improves the performance, in terms of the error, up to 29.89\%, compared to the state-of-the-art solutions.

\end{abstract}


\IEEEpeerreviewmaketitle

\section{Introduction}

Cloud computing environments provide a broad range of advantages for the deployment of scientific applications, especially the ability to provision a large number of resources with a pay-per-use pricing scheme. This characteristic meets the need of scientists that define applications in the form of workflows. Scientific workflows are composed of multiple tasks with dependencies between them. Such workflows are large-scale applications and require a considerable amount of resources to execute. Recent studies show a plethora of algorithms were designed to schedule scientific workflows in clouds \cite{Kousalya2017}. The majority of the solutions are based on heuristic and metaheuristic approaches which attempt to find a near-optimal solution to this NP-hard problem. These optimization techniques in scheduling rely on the estimation of task runtime and resource performance to make scheduling decisions. This estimate is vital, especially in cost-centric, dynamic environment like clouds. An inaccurate estimate of a task's runtime in scientific workflows has a snowball effect that may eventually lead to all of the successors of the task taking longer time than expected to complete. In the end, this will have a negative impact on the total workflow execution time (i.e., makespan) and inflict an additional cost for leasing the cloud resources.

With the emergence of multi-tenant WaaS platforms that deal with a significant amount of data, having a module within the system that can predict the task's runtime in an efficient and low-cost fashion is an ultimate requirement. WaaS is an emerging concept that offers workflow execution as a service to the scientific community. WaaS can be categorized as either Platform as a Service (PaaS) or Software as a Service (SaaS) on the cloud stack service model. WaaS provides a holistic service for scientific workflows execution and deals with the complexity of multi-tenancy, in contrary to a regular WMS that is commonly used for managing the scientific workflow execution of a single user. WaaS platforms are designed to process multiple workflows from different users. In this case, the workload of workflows is expected to arrive continuously, and the workflows must be handled as soon as they arrive due to the quality of service (QoS) constraints defined by the users. Hence, these platforms need to be capable of processing requests in a near real-time fashion. The runtime prediction of tasks must be achieved in a fast and reliable way due to the nature of the environment. Moreover, WaaS platforms make use of the distributed resources provided by the Infrastructure as a Service (IaaS) providers. Therefore, the prediction method should be able to adapt to a variety of IaaS cloud computational infrastructure seamlessly.

Predicting task runtime in clouds is non-trivial, mainly due to the problem in which cloud resources are subject to performance variability \cite{jackson2010performance}. This variability occurs due to several factors--including virtualization overhead, multi-tenancy, geographical distribution, and temporal aspects \cite{Leitner:2016:PCS:2926746.2885497}--that affect not only computational performance but also the communication network used to transfer the input/output data \cite{6848061}. In this area, most of the existing approaches are based on the profiling of tasks using basic statistical description (e.g., mean, standard deviation) to summarize the existing historical data of scientific workflow executions to characterize the tasks, which then is exploited to build a performance model to predict the task runtime. Another approach uses a profiling mechanism that executes a task in a particular type of resource and utilizes the measurement as an estimate. These methods are impractical to adopt in cloud computing environments. Relying only on the profiling based on the statistical description does not capture sudden changes in the cloud's performance. For example, it is not uncommon for a task to have a  longer execution time during a specific time in cloud instances (i.e., peak hours). Hence, averaging the runtime data without considering the temporal factors will only lead to inaccurate predictions. Meanwhile, profiling tasks by executing them in the desired type of resources will lead to an increase in the total execution cost because the profiler requires an extra budget to estimate the runtime.

On the other hand, machine learning approaches can be considered as a state-of-the-art solution for prediction and classification problems \cite{Jordan255}. Machine learning approaches learn the relation between a set of input and its related output through intensive observation from characteristics of the data, usually referred to as features. To capture several aspects that affect the cloud's performance variation, machine learning may provide a better solution by considering temporal changes and other various factors in task's performance as features.

In the case of predictions, the conventional machine learning approaches are based on a regression function that estimates the runtime of a task from a set of features. Evaluating these techniques to predict the task runtime in WaaS platforms is out of our scope. We are interested in exploring various ways of determining the features on which the regression functions depend on. Typical variables that are being used as features to predict the task runtime are based on the workflow application attributes (e.g., input data, parameters) and the specific hardware details (e.g., CPU capacity, memory size) in which the workflows are deployed. This information is relatively easy to extract, and their values are available before the runtime. However, with the rising trend of cloud computing to deploy the scientific workflows, some of these variables that are related to the specific hardware details may become inaccurate to represent the computational capacity due to the performance variability of cloud instances.

Moreover, we found that, as a part of the anomaly detection in executing scientific workflows, some WMS are equipped with the capability to monitor the runtime of tasks by collecting their resource consumptions in a time-series manner. This is a more advanced approach than typical resource consumption monitoring method that store only the single value of the total resource usage of a task's execution. We argue that time-series data of a task's resource consumption may represent better information of a task's execution to be used as features.

Based on these requirements, we propose an online incremental learning approach for task runtime prediction of scientific workflows in cloud computing environments. The online approach can learn as data becomes available through streaming. The online approach is fast since the model only sees and processes a data point once when the task finishes. The incremental approach enables the model to capture the environmental changes such as peak hours in clouds and is capable of adapting to the heterogeneity of different IaaS cloud providers. We also propose to utilize resource monitoring data such as memory consumption and CPU utilization that is collected continuously based on a configurable time interval in the form of time-series records. In summary, the main contributions of this paper are:

\begin{enumerate}
	\item The adoption of online incremental learning approach to predict task runtime in cloud environments.
	\item The use of fine-grained resources monitoring data in the form of time-series records to enhance the task runtime prediction.
\end{enumerate}

The rest of this paper is organized as follows. Section II reviews works that are related to our paper. Section III describes the problem definition. Meanwhile, Section IV explains online incremental learning and Section V describes the proposed solution. Performance evaluation is presented in Section VI. Furthermore, Section VII discusses the results and analysis. Finally, the conclusions and future work are depicted in Section VIII.

\section{Related Work}

The profiling of task's performance in scientific workflows has been extensively studied to create a model that can be used to estimate runtime. The work is useful for improving the scheduling of workflows as the estimation accuracy affects the precision of scheduling algorithms' performance. A study by Juve et al. \cite{JUVE2013682} discusses the characterization and profiling of workflows based on the system usage and requirements that can be used for generating a model for estimating the task runtime based on a basic statistical description. Our work differs in that we use machine learning to predict task runtime instead of the statistical description to summarize the profiling data.

Another work of task runtime prediction for scientific workflows uses an evolutionary programming approach in searching the workflow execution similarities to create a template based on the workflow structure, application type, execution environment, resource state, resource sharing policy, and network \cite{NADEEM2013926}. The template that refers to a set of selected attributes of scientific workflow execution is later used to generate a prediction model for task runtime in a grid computing environment. The use of evolutionary programming is known for its computational intensiveness as the search space increases. It differs from our work which is based on an online approach to achieve fast predictions.

A runtime estimation framework built for ALICE (\textbf{A} \textbf{L}arge \textbf{I}on \textbf{C}ollider \textbf{E}xperiment) profiles a sample of tasks by executing them before the real workflow execution to predict their runtime \cite{PUMMA201765}. The framework captures the features of sample task execution records and uses them as an input for the prediction model. This approach is suitable for massive established computational infrastructures, but may not be appropriate for cloud computing environments. Our work considers clouds, therefore, we avoid additional costs as much as possible by doing extra execution for the sake of profiling to predict task runtime.

\begin{figure}[!t]
	\centering \includegraphics[width=\linewidth]{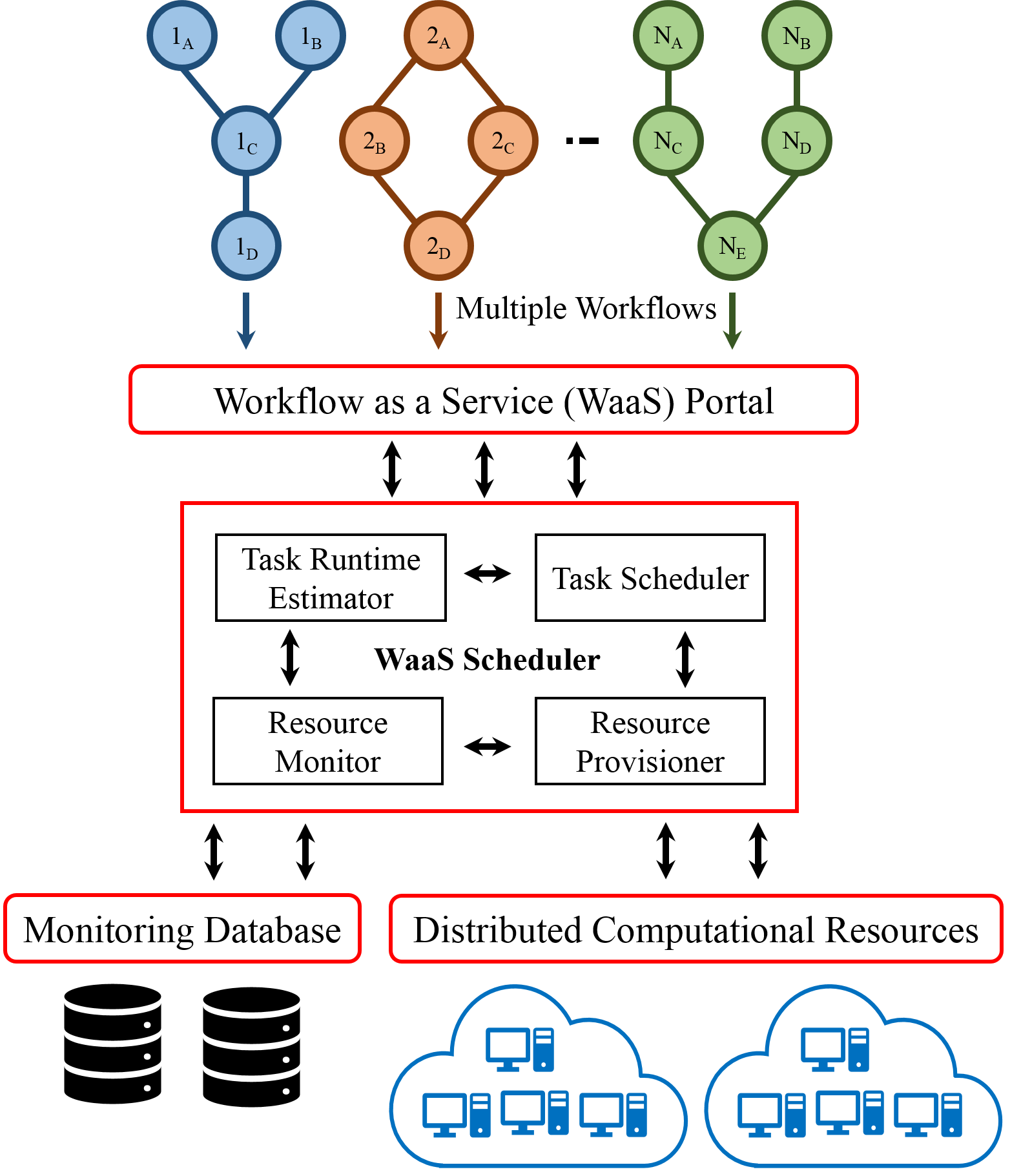}
	\caption{Workflow as a Service (WaaS) architecture}
	\label{figure:archit}
\end{figure}

The works that consider machine learning approaches are dominating the state-of-the-art of task runtime prediction. Regardless of the type of machine learning techniques that are being used, the proposal by Da Silva et al. \cite{doi:10.1142/S0129626415410030} and Pietri et al. \cite{Pietri:2014:PME:2691175.2691178} exploited the workflow application attributes--such as input data, parameters, and workflow structure--as features to build the prediction models. These attributes uniquely adhere to the tasks and are available before the execution. However, in WaaS platforms where the resources are leased from third-party IaaS providers, the hardware heterogeneity may result in different performance even for the same task of the workflow.

In this case, the other works combine the workflow application attributes with specific hardware details where the workflows are deployed, as features. Matsunaga and Fortes \cite{5493447} used application attributes (e.g., input data, application parameters) in combination with the system attributes (e.g., CPU microarchitecture, memory, storage) to build the prediction model for resource consumption and task runtime. Another work by Monge et al. \cite{Monge2015} exploited the task's input data and historical provenance data from several systems to predict task runtime in the gene-expression analysis workflow. The combination of application and hardware attributes provides better profiling of task's execution that arguably results in the improvement of task runtime prediction.

However, only using features for which the values are available before runtime--such as application attributes and hardware details--may not be sufficient to profile the task's execution time in cloud environments. Therefore, further works in this area consider the specific execution environment (e.g., cloud sites, submission time) and the resource consumption status (e.g., CPU utilization, memory, bandwidth) as features. Some of these may be available before runtime (e.g., execution enviroment) but most of them (e.g., resource consumption status) can only be accessed after the task's execution. Hence, the latest variables are mostly known as runtime features as their collection occurs during the task's execution. This runtime features plausibly provide better profiling of the task's execution in cloud environments. The works that exploit this approach such as Seneviratne and Levy \cite{SENEVIRATNE2011245}, used linear regression to estimate this runtime features--such as CPU and disk load--before using them to predict the task runtime. Meanwhile, Pham et al.  \cite{8013738} proposed a similar approach, called two-stage prediction, to estimate the resources consumption (e.g., CPU utilization, memory, storage, bandwidth, I/O) of a task's execution in particular cloud instances before exploiting them for task runtime prediction.

Nonetheless, these related works are based on batch offline machine learning approaches. The batch offline approach poses an inevitable limitation in WaaS platforms. This limitation is related to the streaming nature of workloads in WaaS platforms that need to be processed as soon as they arrive in a near real-time fashion. Our work differs in that we use an online incremental approach and exploit the time-series resource monitoring data to predict the task runtime.

\section{Problem Definition}

This work considers workflows that are modeled as directed acyclic graphs (DAGs), graphs that have directed edges and have no cycles. A workflow $W$ is composed of a set of tasks $T = (t_1, t_2, \dots, t_n)$ and a set of edges $E = (e_{12}, e_{13}, . . ., e_{mn})$ in which an edge $e_{ij}$ represents a dependency between task  $t_i$ that acts as a parent task and task $t_j$ as child task. Hence, $t_j$ will only be ready for execution after $t_i$ has been completed. We assume the execution of these workflows is done via a WaaS platform, and a reference architecture for this system focusing on the scheduler component is shown in Figure \ref{figure:archit}.

The platform consists of a submission portal in which workflows are continuously arriving for execution. These workflows are processed by the scheduler, which is responsible for placing the tasks on either existing or newly acquired resources. The scheduler has four main components: task runtime estimator, task scheduler, resource provisioner, and resource monitor. The task runtime estimator is used to predict the amount of time a task will take to complete in a specific computational resource (i.e., virtual machine). The task scheduler is used to map a task into a selected virtual machine for execution. The resource provisioner is used to acquire and release virtual machines from third-party providers and allocate. The resource monitor is used to collect the resource consumption data of a task executed in a particular virtual machine and the collected data are stored in a monitoring database.

\begin{table}[!t]
	\centering
	\caption{Description of runtime metrics}
	\label{table:profmetricsruntime}
	\resizebox{\linewidth}{!}{\begin{tabular}{@{\extracolsep{4pt}} l l l @{}}
			\hline \noalign{\vskip 1mm}
			\textbf{Resource} & \textbf{Metric}&\textbf{Description} \\
			
			\hline \noalign{\vskip 1mm}
			\multirow{4}{*}{\textbf{CPU}}
			&procs& Number of process\\
			&stime& Time spent in user mode\\
			&threads& Number of threads\\
			&utime& Time spent in kernel mode\\
			\cline{1-3} \noalign{\vskip 1mm}
			\multirow{2}{*}{\textbf{Memory}}&vmRSS& Resident set size\\
			&vmSize& Virtual memory usage\\
			\cline{1-3} \noalign{\vskip 1mm}
			\multirow{7}{*}{\textbf{I/O}}
			&iowait& Time spent waiting on I/O\\
			&rchar& Number of bytes read using any read-like syscall\\
			&read\_bytes& Number of bytes read from disk\\
			&syscr& Number of read-like syscall invocations\\
			&syscw& Number of write-like syscall invocations\\
			&wchar& Number of bytes written using any write-like syscall\\
			&write\_bytes& Number of bytes written to disk\\
			\hline
	\end{tabular}}
\end{table}

In this work, we focus on the task runtime estimator. We assume running tasks are continuously monitored to measure their resource consumption in a specific computational resource. The usage of different resources such as CPU, memory, and I/O are captured by different metrics. These are described in Table \ref{table:profmetricsruntime}. As a result, the data collected for each task and each metric correspond to a series of tuples consisting of a timestamp $t$ and a value $v$ ($<t$, $v>$), where the value corresponds to a specific resource consumption measurement. The measurement's frequency is configurable by a time interval $\tau$. Smaller $\tau$ values translate into more frequent resource consumption measurements, while larger values reduce the frequency and result in less monitoring data. These time-series records are stored in a monitoring database, which are later used by the task runtime estimator. 

We also assume some features describing a given task and its execution environment are available. In particular the task's profile, virtual machine configuration used for the task's execution, and the task's submission time. These are shown in Table \ref{table:profmetrics} and are referred to from now on as pre-runtime features. The problem becomes then on efficiently utilizing these pre-runtime data in conjunction with the resource monitoring time-series data to accurately estimate the runtime of a task in an online incremental manner, that is, as it arrives for execution.

\section{Online Incremental Machine Learning}

In general, machine learning methods are employed to learn some insights from patterns in available data and to predict future events. Classical batch offline learning, which learns from an already collected, and accessible dataset, is not suitable for processing a rapid volume of data in a short amount of time. A reason for this is the fact that these methods usually require the entire dataset to be loaded into memory. Furthermore, batch offline methods do not continuously integrate additional information as the model incrementally learns from new data, but instead reconstruct the entire model from scratch. This is not only time consuming and compute intensive, but also may not be able to capture the temporal dynamic changes in the data statistics. As a result, batch offline learning methods are not appropriate for dynamic environments that introduce and analyze a significant amount of data in a streaming way, such as WaaS platforms.

Instead, online incremental learning has gained significant attention with the rise of big data and internet of things (IoT) trends as it deals with a vast amount of data that does not fit into memory and may come in a streaming fashion. As a result, we propose the use of two algorithms implemented using online incremental learning approaches to estimate the runtime of tasks in a near real-time fashion, namely Recurrent Neural Network (RNN) and K-Nearest Neighbors (KNN). Online incremental learning methods fit naturally into WaaS environments since they incrementally incorporate new insights from new data points into the model and traditionally aim to use minimal processing resources as the algorithms read the new data once available. The other advantage worth noting is that incremental learning enables the model to adapt to different underlying infrastructures. Hence, it enables the creation of models that are agnostic to platforms for deployment.

\begin{table}[!t]
	\centering
	\caption{Description of pre-runtime configuration}
	\label{table:profmetrics}
	\resizebox{.9\linewidth}{!}{\begin{tabular}{@{\extracolsep{4pt}} l l l @{}}
			\hline \noalign{\vskip 1mm}
			& \textbf{Name}&\textbf{Description} \\
			
			\hline \noalign{\vskip 1mm}
			\multirow{3}{*}{\textbf{Task}}
			&name& Name of the task\\			
			&id&ID for a particular type of task\\
			&input&Input name for a task\\
			\cline{1-3} \noalign{\vskip 1mm}
			\multirow{3}{*}{\textbf{VM Type}}
			&memory&Memory capacity\\
			&storage&Storage capacity\\
			&vcpu&Number of virtual processor\\
			\cline{1-3} \noalign{\vskip 1mm}
			\multirow{2}{*}{\textbf{Submission Time}}&day&submission day\\
			&hour&submission hour\\
			\hline
	\end{tabular}}
\end{table}

\subsection{Recurrent Neural Networks}

A special type of RNN called Long Short-Term Memory networks (LSTMs) is capable of remembering information for an extended period of time \cite{Gers99learningto}. Instead of having a simple layer as in regular RNNs, an LSTM network has four unique plus one hidden layers in repeating modules that enable them to learn the context and decide whether the information has to be remembered or forgotten. These layers are a memory unit layer $c$, three types of gate layers- the input gate $i$, the forget gate $f$, and the output gate $o$- plus a hidden state $h$.

For each time step $t$, LSTM receives a set of values $x_t$ corresponding to the different features of the data, and the previously hidden state $h_{t-1}$ that contains the context from previous information as input. Then, LSTM computes the output of the gates based on the activation function which includes the weights and biases of each gates. Finally, this process can be repeated and configured for it to produce an output sequence \{$o_t$, $o_{t+1}$, $o_{t+2}$, $o_{t+3}$, . . ., $o_{t+n}$\} as the prediction result.

Based on these capabilities, LSTM becomes a suitable approach and shows promising results for time-series analysis \cite{Gro2017PredictingTS}. Moreover, it supports online learning since the implementation of LSTM in Keras\footnote{https://keras.io/} provides \emph{batch size} variable that limits the number of data to be trained. A \emph{batch size} value of 1 is used for an online learning approach. Keras also accommodates incremental learning as it incorporates the ability to update the LSTM model whenever new information is obtained continuously.

\subsection{K-Nearest Neighbor}

K-Nearest Neighbor (KNN) is a machine learning algorithm that generates classification/prediction by comparing new problem instances with instances seen in training. KNN computes distances or similarities of new instances to the training instances when predicting a value/class for them. Given a data point $x$, the algorithm computes the distance between that data and the others in the training set. Then, it picks the nearest $K$ training data and determines the prediction result by averaging the output values of these $K$ points.

KNN is widely used for prediction in many areas from signal processing \cite{8316337} to energy management \cite{YESILBUDAK2017434}. It is also used for time-series analysis \cite{Li:2018:1574-8936:14} that resembles sequential problems. One of the implementations of KNN is the IBk algorithm \cite{Aha1991} that is included in the WEKA data mining toolkit \cite{Frank2010}. It incorporates the capability to learn the data incrementally. While the lazy behavior of KNN is compute-intensive and may slow down the performance as the training set increases, IBk incorporates a \emph{window size} variable that enables the algorithm to maintain a number of records from the training set. This capability achieves the trade-off between learning accuracy and speed that is determined by the size of the window by dropping the older data as new data is added to the set. This basic function becomes an advantage for the algorithm to handle the changes in the statistics of the data.

\section{Task Runtime Prediction Using Time-Series Monitoring Data}

In this study, we aim to predict the runtime of the task in a WaaS platform. Given a set of pre-runtime features as listed in Table \ref{table:profmetrics}, we build a model using an online incremental approach that can give an estimate of the time needed to complete a task in a specific virtual machine. In particular, we implement a task runtime estimator module that can be easily plugged in into a WaaS platform and the only requirements being (i) access to the pre-runtime features of a task and (ii) a resource consumption monitoring system that records data in a time-series database. We make use of these data to incrementally build the model as a task finishes its execution. Specifically, when a task is fed into the WaaS scheduler, the algorithm extracts its pre-runtime features and predicts its resource consumption estimation for each metric using LSTM. Then, each resource consumption of a task (i.e., first phase prediction result) is processed to get a representative and distinctive value from the time-series. This process is called feature extraction. Afterward, this value from the feature extraction along with the pre-runtime features are fed to IBk to predict the task's runtime. This process is outlined in Algorithm \ref{algo:onlineprediction}. From now on, we refer to our proposed approach as the time-series scenario.

\begin{algorithm}[!t]
	\caption{Task runtime prediction}\label{algo:onlineprediction}
	\begin{algorithmic}[1]
		\Statex{Input: a task of the workflow $t_i$}
		\Statex{Input: a virtual machine type $v_i$}
		\Statex{Input: submission time $s_i$}
		\Statex{Output: runtime prediction $\alpha$ for $t_i$ on $v_i$ at $s_i$}
		\While{incoming task $t$ in WaaS}
		\Statex\hspace{\algorithmicindent}{\textit{Phase 1:}}
		\State{$\sigma_{i} \gets$ extract pre-runtime features for $t_i$ on $v_i$ at $s_i$}
		\Statex\hspace{\algorithmicindent}{\textit{Phase 2:}}
		\For{selected runtime features $R$ of task $t_i$}
		\State{$\{r_{j_1},r_{j_2},..,r_{j_n}\} \gets$ predict resource}
		\Statex\hspace{\algorithmicindent}\hspace{\algorithmicindent}\hspace{\algorithmicindent}{consumption $r_j$ of $t_i$ using $\sigma_{i}$}
		\State{$\varsigma_j \gets$ extract feature of time-series}
		\Statex\hspace{\algorithmicindent}\hspace{\algorithmicindent}\hspace{\algorithmicindent}{$\{r_{j_1},r_{j_2},..,r_{j_n}\}$ using Equation \ref{equation:feat}}
		\EndFor
		\Statex\hspace{\algorithmicindent}{\textit{Phase 3:}}
		\State{$\alpha \gets$ predict runtime of $t_i$ using $\sigma_i$}
		\Statex\hspace{\algorithmicindent}\hspace{\algorithmicindent}\hspace{\algorithmicindent}{and  a set of features $\{\varsigma_1, \varsigma_2,..,\varsigma_n\}$ from $R$}
		\EndWhile
	\end{algorithmic}
\end{algorithm}

We propose a framework in which multiple prediction models, one for each task in the workflow, are maintained, rather than having a single prediction model for all tasks submitted into the system. We argue that this approach has three main benefits (i) a single prediction model contains information that may act as noise for different tasks, (ii) the size of a single model will grow as the number of tasks increases; this may not be scalable to the size of memory, (iii) multiple models can be maintained by temporarily saving unused models into disk and being loaded whenever the corresponding task needs to be processed. Furthermore, multiple models allow the system to optimize predictions as each model can be fine-tuned to a specific task's requirements (e.g., feature selection).

\begin{figure*}[!t]
	\centering
	\begin{subfigure}{\columnwidth}
		\begin{flushleft}
			\includegraphics[width=.95\columnwidth]{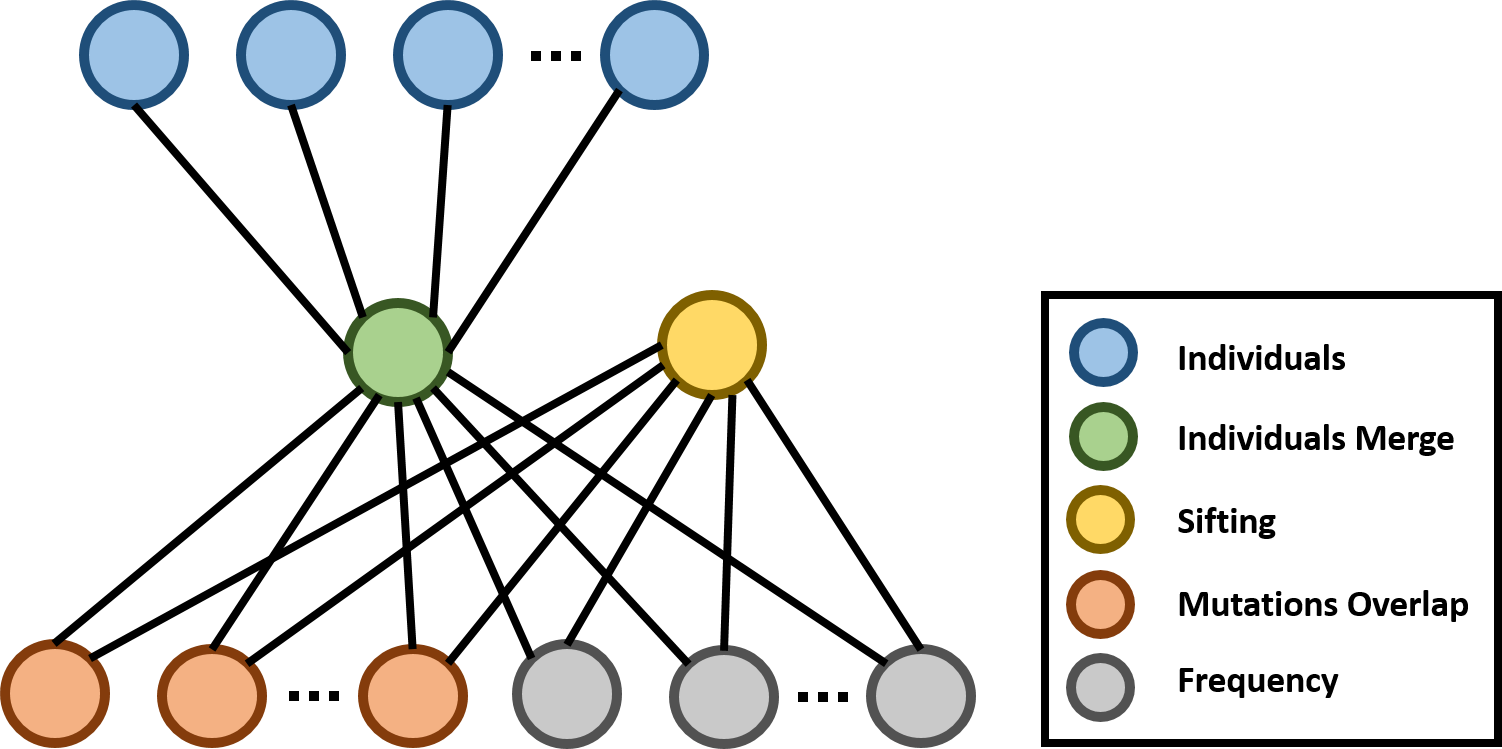}
		\end{flushleft}
		\caption{1000 Genome workflow}
	\end{subfigure}
	\begin{subfigure}{\columnwidth}
		\begin{flushright}
			\includegraphics[width=.85\columnwidth]{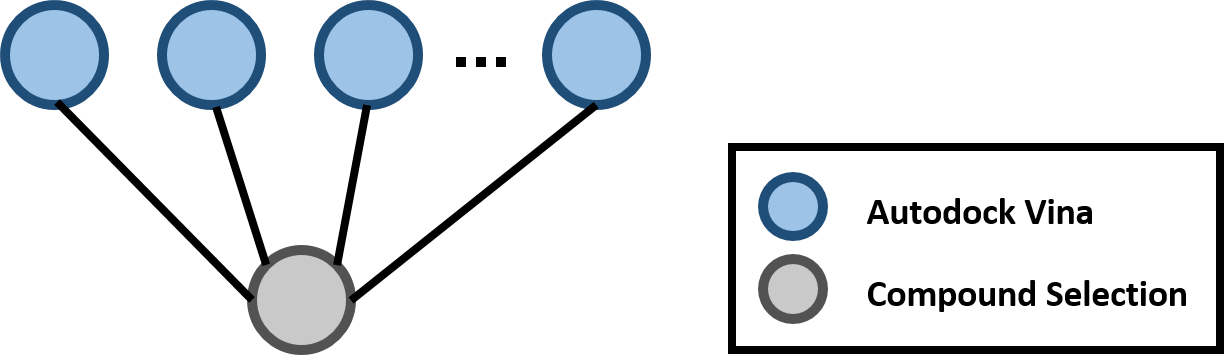}
		\end{flushright}
		\caption{Autodock Vina workflow}
	\end{subfigure}
	\caption{Sample of bioinformatics workflows}
	\label{fig:workflow}
\end{figure*}

To execute the workflows and collect the monitoring data as outlined in Section III, we use the Pegasus \cite{DEELMAN201517} WMS that is equipped with a monitoring plugin as part of the Panorama \cite{doi:10.1177/1094342015594515} project. The monitoring is done at a task level. Therefore, the measurements correspond to the independent execution of a task in a particular type of resource at a specified time.

The first phase of task runtime prediction extracts the pre-runtime configurations $\sigma_i$ of a task $t_i$ and the particular computational resource type $v_i$ where the task will run. These are listed in Table \ref{table:profmetrics}. We decided to include the submission time (i.e., day and hour) to capture performance variability in clouds. For instance, a study by Jackson et al. \cite{jackson2010performance} show that the CPU performance of VMs in clouds was varied by 30\% in compute time. Furthermore, Leitner and Cito \cite{Leitner:2016:PCS:2926746.2885497} suggest that different running time may affect the performance of cloud's resources.

Then, in the second phase, given the set of pre-runtime features $\sigma_i$ of a task $t_i$, we estimate the resource consumption time-series $R_i$ for each metric defined in Table \ref{table:profmetricsruntime} using LSTM. The LSTM model is incrementally updated using data obtained after task $t_h$ finishes executing. These data consist of its pre-runtime features $\sigma_h$ and a set of resource consumption time-series $R_h$ collected during the runtime. LSTM learns the resource consumption sequence per time step $t$ and predicts the value of time step $t+1$ that are separated by time interval $\tau$, and repeats the process until it reaches a desired time-series length $n$ of time step $t+n$. Since every task has a different length of resource consumption record, we padded the end of the sequence with zeros until a specified length and removed the padded values at the end of the prediction. It needs to be noted that not all collected metrics have to be used in the prediction model as features. Feature selection can be done at this stage by calculating the Pearson's correlation coefficient $\rho$ of each metric to the actual task runtime \cite{hall1999correlation}.

The next step in the second phase is time-series feature extraction. The estimated time-series resource consumption $R_i$ for a particular task $t_i$ is pre-processed before being used as feature to estimate the task runtime in the third phase. We used time-reversal asymmetry statistic \cite{6786425} to extract values $\varsigma_i$ from the estimated resource consumption $R_i$ as shown in Equation \ref{equation:feat},
\begin{equation}\label{equation:feat}
\varsigma_i(l)=\frac{\langle(x_{t+l} - x_{t})^3\rangle}{\langle(x_{t+l} - x_{t})^2\rangle^{\frac{3}{2}}}
\end{equation}\\
The feature that is extracted using this algorithm may represent the distinct time-series instance characteristics by calculating a value of specified sub-sequence with a window size determined by lag value $l$ and performing surrogate data test $\langle . \rangle$ across the time-series. In a study by Fulcher and Jones \cite{6786425}, this technique has been proven to be able to classify the time-series dataset of four classes using only one feature without error.

Finally, in the third phase, the extracted relevant features $\varsigma_i$ for a task $t_i$ from the second phase are combined with its pre-runtime features $\sigma_i$ to predict the runtime using the IBk algorithm.

\section{Performance Evaluation}

We evaluate the proposed approach with two workflows from the bioinformatics field. The first workflow is based on the 1000 Genomes Project\footnote{http://www.internationalgenome.org/about}, an international effort to establish a human genetic variation catalog. Specifically, we use an existing 1000 Genome workflow\footnote{https://github.com/pegasus-isi/1000genome-workflow} developed by Pegasus to identify overlapping mutations. It provides a statistical evaluation of potential disease-related mutations. Its structure is shown in Figure \ref{fig:workflow}a. For our experiments, we analyze the data corresponding to three chromosomes (chr20, chr21, and chr22) across five populations: African (AFR), Mixed American (AMR), East Asian (EAS), European (EUR), and South Asian (SAS).

The second workflow uses AutoDock Vina \cite{doi:10.1002/jcc.21334}--a molecular docking application--to screen a large number of ligand libraries for plausible drug candidates (i.e., virtual screening). In particular, we use a virtual screening case of one receptor and forty ligands with various sizes and search spaces of docking box taken from the Open Science Grid Project\footnote{https://confluence.grid.iu.edu/display/CON/Autodock-Vina workflow}. The molecular docking tasks (i.e., AutoDock Vina) in this workflow can be considered as a bag of tasks where every task of receptor-ligand docking can be executed in parallel before the compound selection task takes place to select the drug candidates. The structure of the virtual screening workflows using AutoDock Vina is depicted in Figure \ref{fig:workflow}b.

To the best of our knowledge, this is the first work that predicts the runtime of workflow tasks using an online incremental learning approach. Hence, to compare our work with existing state-of-the-art solutions of task runtime prediction, we reproduce the batch offline learning work by da Silva et al. \cite{doi:10.1142/S0129626415410030} that makes use of a task's input data as a feature to predict the task runtime. We refer to this approach as the baseline scenario. We also replicate the two-stages task runtime prediction in batch offline learning methods by Pham et al. \cite{8013738}, which combined the use of input data, system configuration, and resource consumption to predict task runtime. We refer to this solution as the two-stages scenario. The latest solution is similar to our work except that we use the fine-grained resource consumption time-series data instead of an aggregated value of the consumed resources. We also implement an online incremental version of both solutions to be compared with our proposed approach. To ensure the fairness of each evaluation, we use the IBk algorithm with the default configuration for both batch offline and online incremental learning scenarios.

\subsection{Experiment Setup}

We set up the system on NeCTAR\footnote{https://nectar.org.au/} cloud resources to evaluate the approaches. We use three different virtual machine types from NeCTAR that have the same storage capacity and operating system as depicted in Table \ref{table:vm}. 

\begin{table}[!t]
	\caption{NeCTAR virtual machines configuration}
	\label{table:vm}
	\resizebox{\columnwidth}{!}{\begin{tabular}{@{\extracolsep{4pt}} l c c c c c@{}}
			\hline \noalign{\vskip 1mm}
			\textbf{VM Type} & \textbf{vCPU} & \textbf{Memory} &\textbf{Storage} &\textbf{Operating System} \\
			
			\hline \noalign{\vskip 1mm}
			m2.small&1 & 4GB & 100GB&CentOS 7 (64-bit)\\
			m2.medium&2 & 6GB & 100GB&CentOS 7 (64-bit)\\
			m2.large&4 & 12GB & 100GB&CentOS 7 (64-bit)\\ 
			\hline
	\end{tabular}}
\end{table}

For the experiment, we have generated between 900 and 12,000 executions for every task. The details of these tasks are depicted in Table \ref{table:dataset}. The resource consumption metrics for each running task are collected every time interval $\tau$ seconds, where 1 $\leq$ $\tau$ $\leq$ 30. Specifically, we use $\tau$ values of 1, 5, 10, 15, and 30 to analyze the trade-off between time-series granularity and learning performance. Furthermore, we define the lag values $l$ as $l=2$ and $l=3$ to see the effect of time-series data length on the feature extraction algorithm. However, we do not consider the \emph{sifting} task from the 1000 genome workflow and the \emph{compound selection} task from the virtual screening workflow in our experiments since it has a very short runtime (under 1 second).

Regarding the machine learning algorithms, there are several configurable parameters for each of them. In general, we use the default configurations from their original implementation. It needs to be noted that we do not fine-tune the algorithms to get the optimal configurations for this problem. Hence, further study to analyze the optimal configurations should be done as future work. 

For the LSTM in resource consumption estimation, we use \emph{batch size} $=1$ since the system requires the data is only seen once. Our LSTM implementation uses \emph{sigmoid} as gates activation function, ten \emph{hidden} \emph{layers}, and one hundred \emph{epochs} to train the model. Meanwhile, for IBk, we use the default parameter values used by the version 3.8 of the WEKA library where $k=1$, no distance weighting, and linear function for the nearest neighbors search algorithm. For our batch offline learning experiments, we use various sizes of training data $d$ to see the performance of classical batch offline learning related to the amount of data collection needed for building a good model for prediction. Specifically, we use the $d$ values of 20\%, 40\%, 60\%, and 80\% in the experiments.

To validate the performance of our approach, we use relative absolute error (RAE) as a metric for evaluation as recomended in an empirical study by Armstrong et al. \cite{ARMSTRONG199269} over several alternative metrics as shown in Equation \ref{equation:rae}, 
\begin{equation}\label{equation:rae}
RAE = \frac{\sum_{i=1}^{n} \mid r_{ij} - e_{ij} \mid}{\sum_{i=1}^{n} \mid r_{ij} - \frac{1}{n} \sum_{i=1}^{n} r_{ij} \mid}
\end{equation}\\
where \textit{n} is the number of predictions. The smaller the RAE value, the smaller the difference between the predicted value and the actual observed value.

\begin{table}[!t]
	\centering
	\caption{Summary of datasets}
	\label{table:dataset}
	\resizebox{.95\columnwidth}{!}{\begin{tabular}{@{\extracolsep{4pt}} l l c c @{}}
			\hline \noalign{\vskip 1mm}
			\multirow{2}{*}{\textbf{Workflow}}&\multirow{2}{*}{\textbf{Task Name}} & \multirow{1}{*}{\textbf{Tasks per}} &\textbf{Total Tasks} \\
			&&\textbf{Workflow}&\textbf{Generated}\\
			
			\hline \noalign{\vskip 1mm}
			\multirow{5}{*}{\textbf{1000 Genome}}&individuals & 10 & 9000 \\
			&individuals merge & 1 & 900 \\
			&sifting & 1 & 900 \\
			&mutation overlap & 7 & 6300\\
			&frequency & 7 & 6300\\ \hline \noalign{\vskip 1mm}
			\multirow{2}{*}{\textbf{Virtual Screening}}&autodock vina & 40 & 12000\\
			&compound selection& 1& 3000\\
			\hline
	\end{tabular}}
\end{table}

\begin{table*}[!t]
	\centering
	\caption{Results of task estimation errors (RAE) using online incremental learning approach}
	\label{table:taskerrors}
	\resizebox{.95\textwidth}{!}{\begin{tabular}{@{\extracolsep{4pt}} l r r r r r r r r r r r r@{}}
			\hline \noalign{\vskip 1mm}
			\multicolumn{1}{c}{\multirow{2}{*}{\textbf{Task}}} & \multirow{2}{*}{\textbf{Baseline}} & \multirow{2}{*}{\textbf{Two-Stages}} &\multicolumn{4}{c}{\textbf{Time-Series ($l=2$)}} &\multicolumn{4}{c}{\textbf{Time-Series ($l=3$)}} \\ \cline{4-7} \cline{8-11} \noalign{\vskip 1mm}
			& & &\textbf{$\tau=1s$} &\textbf{$\tau=5s$}&\textbf{$\tau=10s$}&\textbf{$\tau=15s$}&\textbf{$\tau=1s$} &\textbf{$\tau=5s$}&\textbf{$\tau=10s$}&\textbf{$\tau=15s$} \\
			
			\hline \noalign{\vskip 1mm}
			individuals & 64.200 & 57.571&41.748&41.675&41.722&41.175&\textbf{39.180}&40.680&46.601&41.710 \\
			individuals merge & 42.162 & 36.144&34.706&31.474&36.417&42.162&33.300&\textbf{29.553}&42.162&42.162 \\
			mutation overlap & 5.778 & 3.615&\textbf{3.413}&3.682&5.729&5.778&3.861&3.949&5.778&5.778 \\
			frequency & 48.971 & 37.327&35.523&31.039&30.812&32.251&35.386&\textbf{30.499}&35.108&32.368 \\ \hline \noalign{\vskip 1mm}
			autodock vina & 5.380 & 5.153&4.170&4.062&\textbf{4.023}&4.081&4.140&4.045&4.049&4.090 \\
			\hline
	\end{tabular}}
\end{table*}

\begin{table*}[!t]
	\centering
	\caption{Results of task estimation errors (RAE) using batch offline learning approach}
	\label{table:taskerrorsoffline}
	\resizebox{.85\textwidth}{!}{\begin{tabular}{@{\extracolsep{4pt}} l r r r r r r r r r r r@{}}
			\hline \noalign{\vskip 1mm}
			\multicolumn{1}{c}{\multirow{2}{*}{\textbf{Task}}} &\multicolumn{4}{c}{\textbf{Baseline}} &\multicolumn{4}{c}{\textbf{Two-Stages}} \\ \cline{2-5} \cline{6-9} \noalign{\vskip 1mm}
			&\textbf{$d=20\%$} &\textbf{$d=40\%$}&\textbf{$d=60\%$}&\textbf{$d=80\%$}&\textbf{$d=20\%$} &\textbf{$d=40\%$} &\textbf{$d=60\%$}&\textbf{$d=80\%$} \\
			
			\hline \noalign{\vskip 1mm}
			individuals & 65.543 & 62.587&64.523&66.049&59.117&57.625&57.113&\textbf{55.080} \\
			individuals merge & 42.522 & 42.256&37.936&37.424&37.177&35.227&34.312&\textbf{31.129} \\
			mutation overlap & 4.952 & 4.192&4.138&3.967&4.037&3.598&3.188&\textbf{2.936} \\
			frequency & 51.919 & 50.257&49.101&45.740&44.048&39.675&38.493&\textbf{36.622} \\ \hline \noalign{\vskip 1mm}
			autodock vina & 5.036 & 4.820&4.654&4.597&4.847&4.651&\textbf{4.529}&4.627 \\
			\hline
	\end{tabular}}
\end{table*}

\section{Results and Analysis}

In this section, we present and analyze the results of the experiments. We evaluate our proposed approach against the modified online incremental version of the baseline and two-stages scenarios. To ensure the fair comparison, we also present the results of their original batch offline version for task runtime prediction. Furthermore, we discuss the feature selection evaluation for our proposed approach that can improve the performance of the model for predictions.

\subsection{Proposed Approach Evaluation}

We evaluate our proposed approach with various time intervals $\tau$, and time-series lags $l$. The value of time interval $\tau$ affects how often the system records the resource consumption of a particular task and impacts the length of the time-series data. Meanwhile, the value of the lag $l$ that defines the time-reversal asymmetry statistics in feature extraction relies on the length of the time-series. Larger lag values may not be able to capture the distinctive profile of a short resource consumption time-series. Hence, we fine-tune these parameters for each task differently. The results of these experiments are depicted in Table \ref{table:taskerrors}; It is important to note that these do not include the feature selection mechanism in learning as we separate its evaluation in a different section.

\begin{figure}[!b]
	\centering
	\includegraphics[trim=0cm 0cm 9.25cm 0cm, clip=true,width=.65\columnwidth, angle=270]{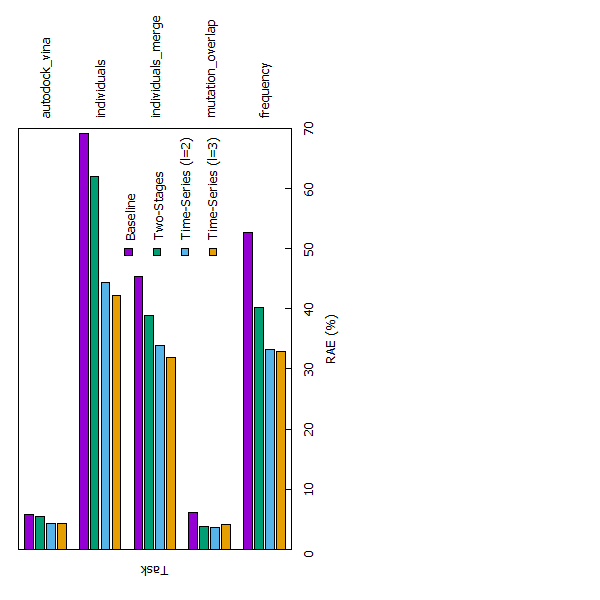}
	\caption{Summary of task estimation errors (RAE) using online incremental learning approach}
	\label{fig:result}
\end{figure}

In general, our proposed approach produces lower RAE compared to the baseline scenario and two-stages scenario. From Figure \ref{fig:result} we can see that exploiting fine-grained resource consumption significantly reduces the RAE of task runtime prediction for \emph{individuals}, \emph{individuals merge}, and \emph{frequency}. Our proposed strategy shows a better result than baseline and two-stages for \emph{mutation overlap} and \emph{autodock vina} although the difference is marginal. In this case, the fine-grained resource consumption features extracted using time-reversal asymmetry statistics may have a higher distinctive property that can characterize each instance uniquely compared to the aggregated value of resource consumption that is being used in the two-stages scenario.

Further analysis from Figure \ref{fig:result} can explain the impact of lag values $l$ on the performance. This graph shows the result from baseline scenario, two-stages scenario, the best result from time-series $(l=2)$ and $(l=3)$ scenarios. As we can see the lag value $l=3$ produces better results than $l=2$ for all cases except \emph{mutation overlap} and \emph{autodock vina} with a marginal difference. In general, a higher lag value means a wider window size of the time-series to be inspected during the time-series feature extraction. However, if the length of the time-series is not long enough, the time-reversal asymmetry statistics cannot fully capture the distinctive characteristics of time-series instance. In \emph{mutation overlap} case, many of the resource consumption time-series length is too short to be evaluated using value $l=3$. Hence, the performance of the algorithm with lag value $l=2$ achieves the lowest RAE. Meanwhile, for \emph{autodock vina}, the trade-off between frequency measurement $\tau$ and lag $l$ cannot be determined as the difference in error results in these various scenarios is insignificant.

Comprehensive results of the online incremental learning approaches can be seen in Table \ref{table:taskerrors}. From the table, we can analyze the impact of configurable parameters to the algorithm's performance. \emph{Individuals} achieves the lowest RAE for $l=3$ and $\tau=1s$. Moreover, \emph{individuals merge} presents the best result for $l=3$ and $\tau=5s$ and \emph{mutation overlap} shows the best result for $l=2$ and $\tau=1s$. Meanwhile, \emph{frequency} gets the lowest RAE for $l=3$ and $\tau=5s$. Lastly, \emph{autodock vina} achieves the lowest RAE for $l=2$ and $\tau=10s$. These results confirm our analysis from the previous discussion related to the length of time-series record and the configurable parameters. Furthermore, from the table, we can see that in several cases, the performance deteriorates to the value of baseline performance. This happens when the time-reversal asymmetry statistics cannot capture the time-series feature property because of the length limitation and the feature extraction algorithm simply gives zero values. Hence, it produces the same result as the baseline scenario.

Therefore, the value of two configurable parameters in time-series feature extraction is an essential aspect in fine-tuning the prediction model. While in general, we can see that a higher lag $l$ value produce a lower RAE, assigning appropriate measurement interval $\tau$ must be further analyzed. There is no exact method to determine this frequency measurement value that is related to the prediction performance. The only known fact that this value inflicts the size of time-series records to be stored in the monitoring database. We leave this problem as future work to improve the task runtime prediction method.

\subsection{Batch Offline Evaluation}

Since the original version of the baseline and two-stage scenarios are implemented in batch offline methods, we also evaluate these approaches to compare with their online incremental version. We use various sizes of training data $d$ and test it using the rest of the data (i.e., 100\% $-$ $d$) for the baseline and two-stages scenarios. The result of this experiments is depicted in Table \ref{table:taskerrorsoffline}. In general, the performance of prediction model improves as the size of data training increases. The results show the same trend for both baseline and two-stages scenarios, but it clearly shows that the two-stages outperform the baseline scenario for all cases. However, the performance of algorithms on more considerable data training becomes a trade-off to the temporal aspect that is critical in WaaS platforms. This criticality is related to the juncture for collecting the data needed to build the model and the speed to compute the data training. Hence, more extensive data training may result in the better algorithm performance but on the other hand, a disadvantage to the WaaS platform. The results show the dependency of batch offline learning methods to the size of data collection for building a prediction model.

\begin{table*}[!t]
	\centering
	\caption{Results of Pearson's correlation based feature selection}
	\label{table:featureselect}
	\resizebox{.75\textwidth}{!}{\begin{tabular}{@{\extracolsep{4pt}} l r r r r r@{}}
			\hline \noalign{\vskip 1mm}
			\textbf{Features}&\textbf{individuals} & \textbf{individuals merge}&\textbf{mutation overlap} &\textbf{frequency}&\textbf{autodock vina} \\
			
			\hline \noalign{\vskip 1mm}
			stime&0.074&0.924&0.435&0.195& 0.570\\
			utime&0.003&0.060&0.995&0.935& 0.974\\ 
			iowait&0.216&0.053&0.006&0.121&-0.008\\
			vmSize&0.027&-0.193&0.518&-0.112&-0.108\\
			vmRSS&0.533&-0.255&0.946&-0.189&-0.129\\
			read\_bytes& 0.004&0.322&0&-0.085&-0.278\\
			write\_bytes& 0.187&-0.463&0.029&-0.237&-0.210\\
			syscr&0.977&-0.608&-0.232&0.130&0.103 \\
			syscw&-0.810&-0.470&-0.153&-0.097&-0.582 \\
			rchar&0.981&-0.490&0.080&0.279&0.071 \\
			wchar&-0.032&-0.454&0.127&-0.212&-0.184 \\
			threads&-0.087&-0.103&-0.408&-0.052&0.019 \\
			procs&-0.087&-0.100&-0.412&-0.052&0.019 \\
			\hline
	\end{tabular}}
\end{table*}

Furthermore, we evaluate our online incremental learning version of the baseline and two-stages scenarios. The results of the online incremental learning approaches are shown in Table \ref{table:taskerrors}. For the baseline and two-stages scenario, the difference in batch offline and online incremental learning is similar in all cases. We notice that the batch offline approaches outperform the online incremental methods in most cases. However, it needs to be noted that such performance is gained after collecting--at least--40\% of the data.

In the end, the improvement of task runtime prediction by using online incremental learning with time-series monitoring data is pretty much significant compared to the conventional batch offline learning methods that rely on the collection of data training beforehand to produce a good prediction model. This criticality limits the batch offline approach to be used in task runtime prediction for WaaS platforms. We argue that both in the case of practicality and performance results, online incremental learning approach using time-series monitoring data may better suit the WaaS environment for task runtime prediction.

\subsection{Feature Selection Evaluation}

Further evaluation is done for the feature selection mechanism. We separate the evaluation to see the real impact of each feature on the learning performance. Hence, we consider the best scenarios from the previous experiment for this evaluation which are time-series scenario with $l=3$ and $\tau=1s$ for \emph{individuals}; $l=3$ and $\tau=5s$ for \emph{individuals merge}; $l=2$ and $\tau=2$ for \emph{mutation overlap}; and $l=3$ and $\tau=5s$ for \emph{frequency}. In Table \ref{table:featureselect} we can see various correlation coefficient values for each feature for each task. A coefficient of zero means the feature is not correlated at all to the task runtime. Meanwhile, a positive correlation value means there is a positive relationship between the feature and the runtime; as the feature value increases or decreases, the runtime follows the same trend. In this case, we select the features with $\abs{\rho}$ values larger than a threshold and evaluate the performance of our approach. There is no exact rule on how to choose the threshold. We choose the value based on small-scale experiments done beforehand, although it needs to be noticed that this value can easily be updated during runtime. Morover, despite various features impacting differently for each task, CPU time (utime and stime), I/O system call (syscr and syscw), and I/O read (rchar) are the most frequent features that exceed the threshold.

\begin{figure}[!b]
	\centering
	\includegraphics[trim=0cm 0cm 10.25cm 0cm, clip=true,width=.6\columnwidth, angle=270]{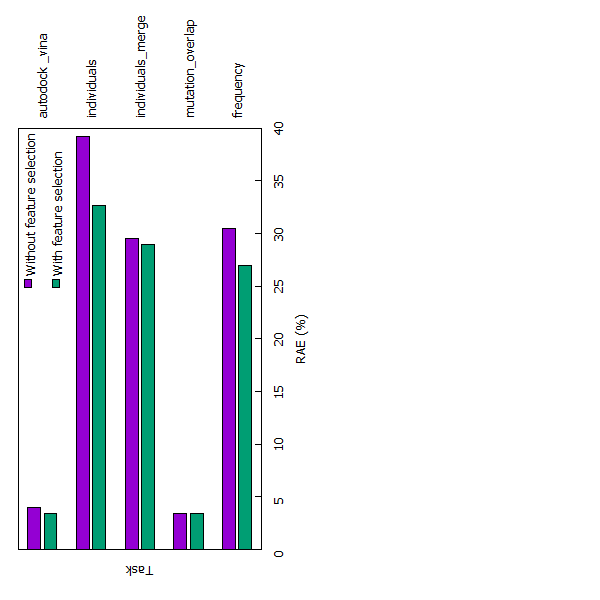}
	\caption{Results of task estimation errors (RAE) with feature selection}
	\label{fig:featureselect}
\end{figure}

From Figure \ref{fig:featureselect} we can see that feature selection impacts the task runtime prediction performance. Significant improvement can be observed for \emph{individuals} and \emph{frequency} with 6.49\% and 3.49\% error reductions respectively.\emph{Individuals merge} show a slightly observed improvement of 0.59\% while the improvement for \emph{mutation overlap} is marginal with 0.04\% error reduction. \emph{Frequency} experiment uses $\abs{\rho}=0.5$ (two features). It only uses a small number of features to outperform the \emph{without feature selection} scenario. Furthermore, \emph{individuals} experiment uses $\abs{\rho}=0.4$ (six features), \emph{mutation overlap} uses $\abs{\rho}=0.09$ (nine features), and the threshold for \emph{individuals merge} is $\abs{\rho}=0.08$ (nine features). The number of selected features are different for each task due to the difference in computational characteristics.

\section{Conclusions and Future Work}

In this paper, we presented online incremental approach for task runtime prediction of scientific workflows in cloud computing environment using time-series monitoring data. The problem of task runtime prediction is modeled based on the requirements for Workflow as a Service (WaaS) platforms which offer the service to execute scientific workflows in cloud computing environments. Hence, approaches to task runtime prediction which use batch offline machine learning may not be suitable in this dynamic environment.

The strategy of using online incremental learning approach is combined with the use of fine-grained resource consumption data in the form of time-series records such as CPU utilization and memory usage. We use a highly distinctive feature extraction technique called time-reversal asymmetry statistics that is capable of capturing the characteristics of a time-series record. Our proposal also considers the selection of features based on Pearson correlation to improve the task runtime prediction and to reduce the computational resources as the system only records the selected relevant features for all tasks.

From our experiments, the proposed approach outperforms baseline scenario and state-of-the-art approaches in task runtime prediction. Although the variation of configurable parameters shows different results, in general, our proposal is better than previous solutions for task runtime prediction. The further result shows that our proposal achieves a best-case and worst-case estimation error of 3.38\% and 32.69\% respectively. These results improve the performance, in terms of error, up to 29.89\% compared to the state-of-the-art strategies.

There are limitations on this work. Further study to evaluate different machine learning algorithms, configurable parameters, and feature selection techniques that best suit task runtime prediction should be carried out. Moreover, the variation of workflow tasks based on their computational characteristics such as data-intensive and compute-intensive tasks should be explored to generate an effective strategy to enhance the performance of the prediction model.

\section*{Acknowledgments}
This research is partially supported by LPDP (Indonesia Endowment Fund for Education) and ARC (Australia Research Council) research grant.

\balance

\bibliographystyle{IEEETran}

\bibliography{mybibfile}

\end{document}